# AN INTERNET REVIEWS TOPIC HIERARCHY MINING METHOD BASED ON MODIFIED CONTINUOUS RENORMALIZATION PROCEDURE

LIN QI*,¶, FEI-YAN GUO*,†, JIAN ZHANG*,‡,¶, YU-WEI WANG§

*School of Economics and Management

Beijing Information Science and Technology University, Beijing 100192, P.R.China

†Beijing World Urban Circular Economy System (Industry) Collaborative Innovation Center,

Beijing 100192, P.R.China.

‡Beijing International Science and Technology Cooperation Base of Intelligent Decision and Big

Data Application, Beijing 100192, P.R.China.

§Department of East Asian Studies, University of Arizona, Tucson AZ85719, U.S.A.

¶qilin@bistu.edu.cn, zhangjian@bistu.edu.cn

## Abstract

Mining the hierarchical structure of Internet review topics and realising a fine classification of review texts can help alleviate users' information overload. However, existing hierarchical topic classification methods primarily rely on external corpora and human intervention. This study proposes a Modified Continuous Renormalization (MCR) procedure that acts on the keyword co-occurrence network with fractal characteristics to achieve the topic hierarchy mining. First, the fractal characteristics in the keyword co-occurrence network of Internet review text are identified using a box-covering algorithm for the first time. Then, the MCR algorithm established on the edge adjacency entropy and the box distance is proposed to obtain the topic hierarchy in the keyword co-occurrence network. Verification data from the Dangdang.com book reviews shows that the MCR constructs topic hierarchies with greater coherence and independence than the HLDA and the Louvain algorithms. Finally, reliable review text classification is achieved using the MCR extended bottom-level topic categories. The accuracy rate (P), recall rate (R) and F1 value of Internet review text classification obtained from the MCR-based topic hierarchy are significantly improved compared with four target text classification algorithms.

## 1. INTRODUCTION

Internet reviews are texts with certain logical structures published by individuals from various perspectives based on their subjective impressions of products or services.[1] The review has various forms, contains fragmented content and has a low information density. Extracting text topics from massive review texts and performing effective topic classification, are helpful to quickly analyze the content of reviews and efficiently mine information, which is convenient for consumers to make decisions and businesses to determine needs.[2] Therefore, the current requirements of various users for a comprehensive and fine-grained understanding of review texts continue to improve. The topic classification results are required to reflect the overall characteristics and specific attributes of products and services. Constructing the topic hierarchy of review texts and achieving the precise and detailed text classification has become the basis for supporting users in understanding product and service information comprehensively, accurately and quickly.[3, 4]

There are more non-hierarchical algorithms and fewer hierarchical algorithms in existing topic mining and subsequent text classification. In non-hierarchical algorithms, the typical unsupervised classification method is LDA, but it is less effective at classifying short text.[5, 6] The typical supervised classification method is TextCNN,[7] but the model needs to be trained in advance, which has a large workload and poor portability. There are two main types of hierarchical algorithms: those based on probabilistic topic models and those based on frequent term mining. Hierarchical Latent Dirichlet Allocation (HLDA)[8] is the most commonly used probabilistic topic model method, but the topic hierarchy obtained has a lot of topic overlap phenomena. The hierarchical structure and topic discrimination are tough to control, and the result's readability is difficult to ensure. The quality of frequent item phrases is extremely important in the method based on frequent item mining.[9, 10] It needs to rely on external information, such as the ontology corpus, which is unsuitable for colloquial Internet review texts. Furthermore, after transforming the text into a keyword co-occurrence network, hierarchical topic mining can be performed according to the hierarchical communities in the complex network. The Louvain algorithm[11] is a hierarchical community detection algorithm with superior time performance. Nevertheless, the coherence of the topics obtained at the same layer is poor, which affects the performance of subsequent text classification. Therefore, it is necessary to propose a suitable model for the topic hierarchy mining of Internet reviews.

Fractal is a structure with self-similarity characteristics. The components of this structure are similar in a way to the whole, displaying similar and fine hierarchical characteristics.[12] The fractal structure widely exists in various complex self-organised systems. For example, at the microscopic scale, fractal growth in the case of crystalline phase transitions,[13, 14] the structure of protein molecules,[15] etc. At the mesoscale, fractal structures exist in human social networks,[16] the urban system,[17, 18] the cauliflower flower ball[19] and other plants. At the macroscale, the appearance and trend of mountains and rivers[20] and structure of infra-red cirrus in the cosmos[21] also have fractal features.

Language is the tool for human communication in the social system. Driven by daily use, the language system constantly derives and develops to improve its complexity and refinement.[22, 23] Therefore, the language system should be a dynamic and self-organising complex system with fractal properties. As a dynamic natural language, review texts express topic granularity from generalisation to specificity, gradually decomposed and refined, so that high-frequency topics are located at the upper level to summarise the evaluation objects, and low-frequency topics are located at the lower level to reflect the detailed evaluation of specific attributes, which are performed in a hierarchical form, refined, supplemented and similar at all levels, converging to the 'labour-saving principle'.[1, 24] Therefore, identifying the fractal characteristics in the review text structure makes it possible to effectively extract the topic hierarchy of reviews through fractal theory.

The advancement of fractal networks research enables the construction of a topic hierarchy of review texts based on natural language processing. As the core feature of the fractal structure, self-similarity is identified by the scale-invariant renormalization procedure. Song et al. proposed a box-covering method to complete the renormalization procedure, which recognised fractal and self-similarity properties in complex networks for the network structure.[25] These boxes often correspond to different functional modules in the network and encode the information about the corresponding modules.[26] The research on fractal networks and the renormalization procedure has been applied in many fields such as rail transit, mobile social and so on, showing the potential of fractal networks and the renormalization procedure.[27-30]

Hence, to make full use of the fractal characteristics in the review text to extract the topic hierarchy, this study proposes a Modified Continuous Renormalization (MCR) procedure that acts on the keyword co-occurrence network. The fractal characteristics in the keyword co-occurrence network of Internet review text are identified using a box-covering algorithm. Then, the MCR algorithm established on the edge adjacency entropy and the box distance is proposed to obtain the topic hierarchy in the co-occurrence network. Finally, the reliable review texts classification is achieved using the MCR extended bottom-level topic categories. The rest of the study is organised as follows: In Sec. 2, the cited theories and algorithms are introduced. Sec. 3 explains the identification method of the topic hierarchy of review texts and the text classification model. In Sec. 4, the Dangdang.com review texts are used to verify the method. The performance comparison with related methods is discussed in Sec. 5. Finally, in Sec. 6, research conclusions and limitations are given.

## 2. RELATED THEORIES AND ALGORITHMS

### 2.1. TextRank Algorithm

The TextRank algorithm is a text keyword extraction method. It takes the words extracted from the text as network nodes, constructs a graph model of words according to the co-occurrence situation, and calculates the importance of words through iteration on the graph to identify keywords.[31] The algorithm can be expressed as a weighted graph model $G = (V,\ E,\ W)$, where

$V$ is the set of word nodes in the graph, $E$ is the set of edges, and $W$ is the set of edge weights. The TextRank value $S(v_i)$ of the word node is defined as follows:

$$S(v_i) = (1 - d) + d \sum_{v_j \in in(v_i)} \frac{w_{ij}}{\sum_{w_k \in out(v_j)} w_{jk}} S(v_j) \qquad (1)$$

Where $in(v_i)$ is the set of nodes pointing to the word node $v_i$; $out(v_i)$ is the set of nodes pointed to by the word node $v_i$; $w_{ij}$ is the edge weight between $v_i$ and $v_j$; $d$ is the damping coefficient, which is generally taken as 0.85.

2.2. Keyword Co-occurrence Network

The keyword co-occurrence network is a complex network composed of textual keyword co-occurrence relationships, which is used to investigate the relationship between topic categories in a particular domain.[32] Keyword co-occurrence refers to the existence of a pair of keywords $\{k_i, k_j\}$ in the candidate keyword set $K = \{k_1, k_2, \ldots, k_n\}$ of the text set. If $k_i, k_j$ co-appears in a text, it is considered that $k_i, k_j$ co-occur. Keyword co-occurrence network $G = (V, E, W)$ is an undirected weighted network, where the node set $V = \{v_i \mid i = 1, 2, \cdots, n\}$ is determined by the text keywords, the edge set $E = \{e_{ij} \mid i, j = 1, 2, \cdots, n\}$ is the co-occurrence relationship between keywords and the edge weight set $W = \{w_{ij} \mid i, j = 1, 2, \cdots, n\}$ is determined by the co-occurrence frequency of keywords. The higher the frequency, the greater the weight, suggesting that the topics represented by the keywords are more closely related.

2.3. Sampling-based Box-covering Algorithm

The box-covering algorithm is used to identify the fractal and self-similarity properties of the network. In the box-covering algorithm, given a box size $\ell_B$, the distance between any two nodes in the box must be less than $\ell_B$, and the rule is to tile the entire network with the minimum number of boxes $N_B$.[25] If there is a power-law relationship between $N_B$ and $\ell_B$:

$$N_B \sim \ell_B^{-d_B} \tag{2}$$

The network shows fractal characteristics, and $d_B$ is the box-covering dimension of the network.

After the entire network is covered by boxes, a box becomes a node and all edges between the original nodes in different boxes degenerate into an edge between the new nodes, which is called the renormalization procedure of the network. The network can be continuously renormalized using boxes of constant size until it degenerates into a node, called the Continuous Renormalization (CR) procedure of the network. As shown in Fig. 1, each column represents the use of boxes of different sizes to cover the network to identify its fractal characteristics. Each row represents the continuous coverage of the network with a constant box size, that is, the CR procedure of the network.

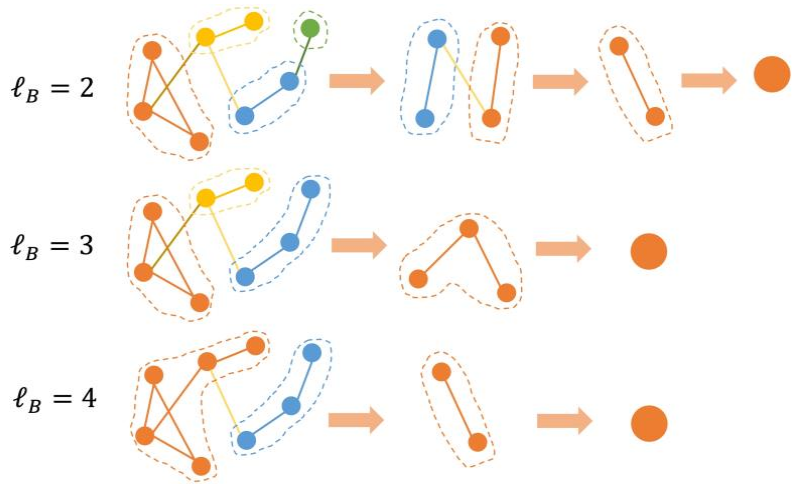

Fig. 1 The renormalization procedure of complex networks.

However, given a constant box size, it is an NP-hard problem to cover a network with the minimum box amount.[25] Therefore, the sampling-based box-covering algorithm is used to improve the solution's efficiency.[33] The 'maximal box sampling' strategy is selected to generate boxes, and the 'big-box-first' greedy algorithm is chosen to select boxes so that the algorithm can complete the solution within an acceptable time, as shown in Algorithms 1 and 2.[33]

Algorithm 1 The maximal box sampling method

| |
|---|
| Input: $G = (V, E)$, $\ell_B$, $n$ |
| Output: $Boxes$ |
| Start |
| 1      $Boxes \leftarrow [\ ]$; $U \leftarrow V$ |
| 2      for $i = 1$ to $n$ do |
| 3         if $U \neq \emptyset$ then |
| 4            $u \leftarrow$ select a node from $U$ randomly |
| 5         else |
| 6            $u \leftarrow$ select a node from $V$ randomly |
| 7         end if |
| 8         $box \leftarrow \{u\}$ |
| 9         $R \leftarrow$ select all the nodes in $V$ whose distance from $u$ is less than $\ell_B$ |
| 10       while $R \neq \emptyset$ do |
| 11           $t \leftarrow$ select a node from $R$ randomly |
| 12           add $t$ to $box$ |
| 13           remove $t$ from $R$ |
| 14           remove all the nodes in $R$ whose distance from $t$ is not less than $\ell_B$ |
| 15       end while |
| 16       append $box$ to $Boxes$ |
| 17       remove all nodes contained in $box$ from $U$ |
| 18     end for |
| End |

Algorithm 2 The big-box first strategy

| |
|---|
| Input: $Boxes = [b_i]$, $S = [\ s_i = \|b_i\|\ ]$ |
| Output: $Box$ |
| Start |
| 1      $U \leftarrow V$, $Box \leftarrow$ all-zero array of length $\|V\|$, $c \leftarrow 0$ |
| 2      sort $Boxes$ ascending by corresponding $S$ values |
| 3      while $s_j > 0$ with $s_j$ being the size of the last element of $Boxes$ do |
| 4         $c \leftarrow c + 1$ |
| 5         $n \leftarrow b_j \cap U$ |
| 6         $Box(n) \leftarrow c$ |
| 7         remove all the nodes contained in $b_j$ from $U$ |
| 8         $S \leftarrow [\ s_i = \|b_i \cap U\|\ ]$ |
| 9         sort $Boxes$ ascending by corresponding $S$ values |
| 10     end while |
| End |

2.4. Text Classification Method Based on Probabilistic Topic Model

The probability topic model text classification method is a text classification method that analyses the text's topic probability distribution. The text is represented as a three-level probability model of 'text-topic-words' in the probabilistic topic model.[5] The topic is represented as a probability distribution of multiple word, and the same word may belong to multiple topic with different attribution probabilities. The text is composed of words, and a certain topic generates each word. Therefore, the text is represented as the probability distribution of multiple topic, and the topic with the highest probability is selected as the category of the text to realise the classification of the text.

Suppose that the topic $z_k = \{p_{v_1}^k, p_{v_2}^k, \cdots, p_{v_m}^k,\}$ consists of $m$ keywords and the probability that the keyword $v_i$ belongs to the topic $z_k$ is $p_{v_i}^k = 1/m$. The discriminant formulas of the text category are as follows:

$$p(y \in z_k) = \frac{\sum_{x_j \in z_k} p_{x_j}^k \cdot td_{x_j}^y \ (j = 1,2,\cdots,n)}{\sum_{k=1}^{K} \sum_{x_j \in z_k} p_{x_j}^k \cdot td_{x_j}^y \ (j = 1,2,\cdots,n)} \quad (3)$$

$$z_{result} = \max(p(y \in z_k)) \ (k = 1,2,\cdots,K) \quad (4)$$

Where $p(y \in z_k)$ is the probability that text $y$ belongs to the topic $z_k$; $z_{result}$ is the topic category corresponding to the highest probability, that is, the final category of text $y$. $n$ is the number of keywords contained in text $y$; $K$ is the number of topics; $x_j$ is the keyword in text $y$; $p_{x_j}^k$ is the distribution probability of keyword $x_j$ under topic $z_k$; $td_{x_j}^y$ is the TF-IDF value of $x_j$ under text $y$.[34]

## 3. TOPIC HIERARCHY MINING AND TEXT CLASSIFICATION MODEL

The overall research framework in this study is shown in Fig. 2, consisting of three stages: (i) Identify the fractal characteristics of the review text keyword co-occurrence network. After preprocessing the review text, keywords are extracted using the TextRank algorithm, the keyword co-occurrence network is then built, and the network fractal characteristic is identified through box-covering. (ii) Identification of the topic hierarchy. Box-covering is performed on the keyword co-occurrence network, and the box distance and edge adjacency entropy are then calculated. The MCR procedure is implemented to obtain the fine topic hierarchy of review texts. (iii) Review text classification. The bottom-level topic features are extended, and the topic's probability distribution of the review text is calculated to achieve the review text classification.

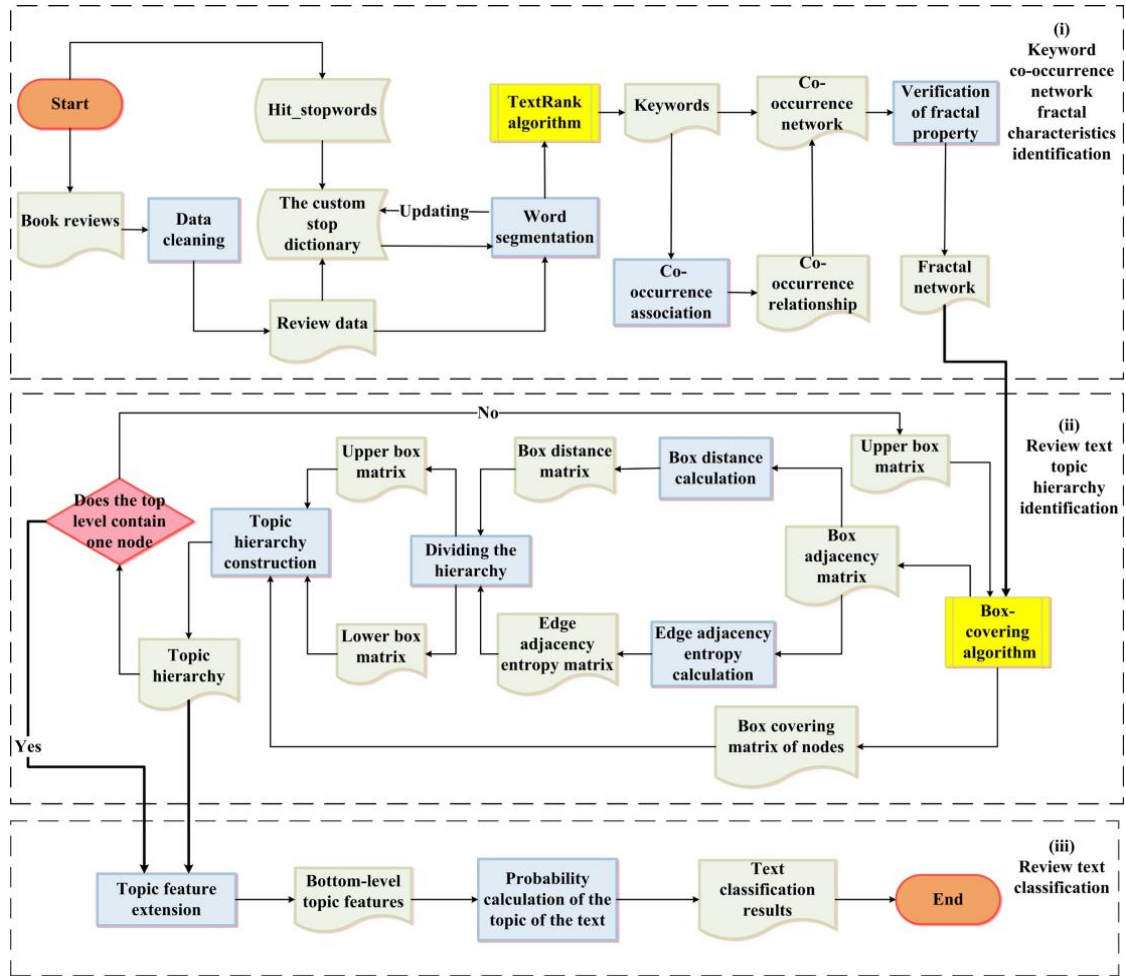

Fig. 2 Research framework.

3.1. Keyword Co-occurrence Network Fractal Characteristics Identification

3.1.1. Data pre-process

The review data are mixed and irregular, which is unsuitable for direct application and needs data preprocessing. The specific steps are as follows: (i) Data cleaning. Duplicate and empty data lines are deleted. The traditional Chinese is also converted to simplified Chinese to unify the word format. (ii) Chinese word segmentation. The pkuseg[35] is used for Chinese word segmentation. (iii) Removing stop words. After Chinese word segmentation, there will be many meaningless words in the construction of the topic hierarchy, which must be filtered out. The

HIT_stopwords dictionary and a custom stop dictionary are used to remove meaningless words in this study.

3.1.2. Fractal characteristics identification

The TextRank algorithm in Sec. 2.1 is used to extract keywords, and then the keyword co-occurrence network is constructed according to Sec. 2.2. Since the box-covering algorithm of the weighted network is still immature, there is only one BCANw algorithm,[36] which is based on the graph colouring problem, and the graph colouring problem itself is an NP-hard problem.[37] Therefore, the weighted co-occurrence network is converted into an unweighted network so that a box-covering algorithm suitable for the unweighted network can be subsequently used. The edge weight of the co-occurrence network reflects the co-occurrence frequency of the keywords. To reflect the influence of the weight, and the medium and high-frequency words that can better reflect the topic reserved, the edges with edge weights less than $\alpha$ are removed. In the experiments in Sec. 4.2, the edge weight threshold $\alpha = 2$ is set, which can retain the most edges and nodes to meet the comprehensiveness requirements of the hierarchical topics. Furthermore, the network used for box-covering needs to be a connected graph, so the nodes with the degree of 0 after deleting edges are removed, and the weight of the remaining edges is set to 1, thereby obtaining an undirected and unweight keyword co-occurrence network. Then, the network is covered with boxes of different scales according to Sec. 2.3 to identify its fractal characteristics.

3.2. Review Text Topic Hierarchy Identification

In the applications of fractal networks and their renormalization, it is necessary to modify the box-covering or renormalization algorithm to adapt to the characteristics of the research problems.[27, 38, 39] The literature shows that in text mining, intermediate frequency words are

keywords that contain more topic meanings, which can better reflect the topic features of the text.[40] Due to the self-reinforcement in the classic box-covering process, the boxes containing high-frequency words are likely to merge with boxes containing intermediate frequency words continuously. Therefore, the edge adjacency entropy and box distance are defined to help construct the MCR algorithm to effectively identify the topic hierarchy.

3.2.1. Edge adjacency

Edge adjacency entropy $h(x)$ describes the dispersion of the edges of a box $x$. It is used to divide a hierarchy of boxes. Calculated as follows:

$$p_x(y) = N(x,y)/s(x) \tag{5}$$

$$h(x) = -\sum_{p_x(y)\neq 0} p_x(y)\log p_x(y) \tag{6}$$

Where $N$ represents the box adjacency matrix; $N(x,y)$ represents the number of edges between $x$ and $y$, which value is 1 or 0; $s(x)$ represents the number of boxes connected by $x$; $p_x(y)$ represents the edge normalisation value of $x$.

Topic boxes with larger edge adjacency entropy have relatively scattered edges and relatively vague meanings, and they should be located at a higher and coarser level. Topic boxes with smaller edge adjacency entropy have relatively concentrated edges and relatively cohesive meanings, and they should be located at a lower and more refined level.

3.2.2. Box distance

The box distance $D_{xy}$ describes the dispersion of boxes $x$ and $y$. It is used for topic association at different levels. Calculated as follows:

$$D_{xy} = \frac{\sum_{i\in b_x, j\in b_y} d_{ij}}{|b_x|\cdot|b_y|} \tag{7}$$

Where $d_{ij}$ represents the distance between node $i$ and node $j$; $b_x$ represents the set of nodes covered by box $x$; $|b_x|$ represents the number of nodes covered by box $x$.

A small box distance means that the node's dispersion between boxes is small. Therefore, a lower-level box should be associated with the higher-level box which has the smallest box distance from it to reflect the relevance in the topic hierarchy.

3.2.3. Topic hierarchy identification of review texts based on the MCR algorithm

Based on definitions of edge adjacency entropy and box distance, the Modified Continuous Renormalization (MCR) algorithm is proposed to identify the topic hierarchy of review texts. As shown in Algorithm 3, there are three main processes: (i) Cover the network with boxes so that nodes are at the lower level of the topic hierarchy, and the corresponding topic boxes will be at the upper level. To make the topic clear and focused, remove boxes with node counts less than $\beta$. This is due to the general 'edge problem' of the current box covering algorithms, that is, many small boxes with few nodes and inconspicuous features exist at the edge of the network, rather than being evenly distributed throughout the network.[41, 42] The features contained in such boxes are generally already included in the selected large boxes. Still, these boxes are left over due to the inherent characteristics of the algorithm so they can be removed. In the experiments in Sec. 4.3, the threshold β = 2 is set, which can remove the edge box with the smallest scale to make the topic features more obvious and meet the fine-grained requirements of hierarchical topics. (ii) According to Eqs. (5)–(7), calculate the edge adjacency entropy and box distance. Topic boxes obtained in (i) are further divided into layers so that boxes with relatively smaller edge adjacency entropy are located in the lower layer of the topic hierarchy, and boxes with larger edge adjacency entropy are located in the upper layers. (iii) If there is more than one topic box at the top level, repeat the processes of (i) and (ii) until the root of the topic hierarchy is obtained and the topic hierarchy identification of the review text is completed.

Algorithm 3  Modified Continuous Renormalization（MCR）

| |
|---|
| Input: $M$; $N$; $n$; $\ell_B$ |
| Output: $H\_Topics$; $S\_Boxes$ |
| Start |
| 1  $runs \leftarrow 1; H\_Topics \leftarrow [\ ]; S\_Boxes \leftarrow [\ ]; H \leftarrow [\ ]\ ; D\_Box \leftarrow [\ ]\ ; U \leftarrow [\ ]; L \leftarrow [\ ]$ |
| 2  while $|N| \neq 1$ |
| 3    $[Node\_Box, Net\_Box, Box] \leftarrow BM(M, n, \ell_B)$ |
| 4    if $runs = 1$ then |
| 5      for $b_x$ in $Node\_Box$ do |
| 6        if the number of nodes contained in $b_x < \beta$ |
| 7          remove the nodes contained in $b_x$ from $N$ |
| 8          remove $b_x$ from the $Node\_Box$ and $Net\_Box$ |
| 9          add $b_x$ to $S\_Boxes$ |
| 10        end if |
| 11      end for |
| 12    end if |
| 13    add $N$ to $H\_Topics$ |
| 14    for $b_x$ in $Node\_Box$ do |
| 15      $h(x) \leftarrow$ the edge adjacency entropy of $b_x$ |
| 16      add $h(x)$ to $H$ |
| 17    end for |
| 18    for $b_x\ b_y$ in $Node\_Box$ do |
| 19      $D_{xy} \leftarrow$ the box distance of $b_x, b_y$ |
| 20      $D\_Box(x, y) \leftarrow D_{xy}$ |
| 21    end for |
| 22    $m \leftarrow$ the maximum value in $H$ plus 1 |
| 23    while the values in $H$ are not equal do |
| 24      $b_x \leftarrow$ the box with the minimum $h(x)$ |
| 25      $b_y \leftarrow$ the box with the smallest $D_{xy}$ from $b_x$ |
| 26      add $b_y$ to $U$;  add $b_x$ to $L$ |
| 27      $h(x), h(y) \leftarrow m$ |
| 28    end while |
| 29    $U\_Net \leftarrow Net\_Box(U); L\_Net \leftarrow Net\_Box(L)$ |
| 30    add $L, U$ to $H\_Topics$ |
| 31    $M \leftarrow U\_Net; N \leftarrow U; runs \leftarrow runs + 1$ |
| 32  end while |
| End |

3.3. Review Text Classification

Based on the topic categories obtained from the MCR algorithm, the classification of review texts is realised. The process is divided into two stages: (i) Use the bottom-level topic expansion algorithm shown in algorithm 4 to classify topic boxes deleted during the operation of the MCR algorithm into bottom-level topic boxes with the closest box distance from them and the smaller edge adjacency entropy. (ii) Use the bottom-level topic categories obtained from algorithm 4, the classification of the review texts is completed according to the text classification method described in Sec. 2.4.

Algorithm 4 Bottom-level Topic Expansion

| |
|---|
| Input: $S\_Boxes = [b_x]$; $Bottom\_Boxes = [b_y]$; $C = [c_i]$ |
| Output: $S\_Boxes = [b_x]$ |
| Start |
| 1   $Z \leftarrow [\ ]; H \leftarrow [\ ] ; D\_Box \leftarrow [\ ] ; Z_{result} \leftarrow [\ ]$ |
| 2   for $b_y$ in $Bottom\_Boxes$ do |
| 3       $h_y \leftarrow$ the edge adjacency entropy of $b_y$ |
| 4       add $h_y$ to $H$ |
| 5   end for |
| 6   for $b_x$ in $S\_Boxes$, $b_y$ in $Bottom\_Boxes$ do |
| 7       $D_{xy} \leftarrow$ the box distance of $b_x$, $b_y$ |
| 8       add $D_{xy}$ to $D\_Box$ |
| 9   end for |
| 10  for $b_x$ in $S\_Boxes$ do |
| 11      $b_y\_set \leftarrow$ the boxes with the smallest $D_{xy}$ from $b_x$ |
| 12      $b_{opt\_y} \leftarrow$ the box with the smallest $h_y$ in $b_y\_set$ |
| 13      expand $Bottom_{Boxes}$ according to $b_{opt\_y} \leftarrow b_{opt\_y} \cup b_x$ |
| 14  end for |
| End |

## 4. EXPERIMENTS AND RESULTS

4.1. Datasource

The purchase reviews of the Top 10 best-selling books on Dangdang.com in China for the past four years (2018–2021) are the data source for the experiments. Totally 275,945 purchase reviews of 18 books in five categories were collected, including novel, literature, encouragement, biography and history, as shown in Table 1. The Dangdang.com book reviews were chosen as the data source since (i) Book review data is typical review data, and the expressed topics may have hierarchical characteristics about the shopping experience, reading experience, harvest evaluation and so forth; (ii) The data amount is appropriate. After preprocessing, the dataset of a book generally contains 10,000 to 30,000 pieces; (iii) The data are easy to obtain using the technical means of the web crawler.

Table 1 Purchase Reviews of 18 Books in Five Categories.

| Category | Book title | Number of reviews (items) | Category | Book title | Number of reviews (items) |
|---|---|---|---|---|---|
| Novel | *No Longer Human* | 16788 | Novel | *Sherlock Holmes* | 19666 |
| | *The Kite Runner* | 6128 | Literature | *Six Records of a Floating Life* | 13758 |
| | *The Little Prince* | 21068 | | *The Complete Works of Echo* | 7712 |
| | *The Moon and Six Pence* | 18032 | | *Poetic Remarks on the Human World* | 16437 |
| | *The Three-Body Problem* | 16028 | Encouragement | *Simplifying Life* | 14491 |
| | *World of Plainness* | 6193 | | *The Home Letter from Zeng Guofan* | 24946 |
| | *The Scholars* | 13792 | Biography | *Decisive Moments in History* | 20327 |
| | *All Men are Brothers* | 27373 | | *Zeng Guofan* | 6636 |
| | *The Complete Works of Eileen Chang* | 11476 | History | *A Brief History of Humankind* | 15094 |

Table 2  Typical Topological Features of Keyword Co-occurrence Networks.

| Id | Typical book | N | E | <k> | D | C | M | $d_B$ |
|---|---|---|---|---|---|---|---|---|
| 1 | *The Home Letter from Zeng Guofan* | 622 | 2993 | 9.624 | 7 | 0.719 | 0.429 | 4.656 |
| 2 | *No Longer Human* | 535 | 2174 | 8.127 | 7 | 0.714 | 0.495 | 4.775 |
| 3 | *Six Records of a Floating Life* | 426 | 1595 | 7.488 | 8 | 0.656 | 0.550 | 3.853 |
| 4 | *Decisive Moments in History* | 488 | 1595 | 6.537 | 8 | 0.656 | 0.562 | 4.940 |
| 5 | *A Brief History of Humankind* | 366 | 1413 | 7.721 | 7 | 0.613 | 0.391 | 4.006 |
| 6 | *The Kite Runner* | 144 | 365 | 5.069 | 7 | 0.545 | 0.575 | 3.361 |
| 7 | *The Moon and Six Pence* | 467 | 1903 | 8.150 | 7 | 0.703 | 0.447 | 4.459 |
| 8 | *The Three-Body Problem* | 452 | 1709 | 7.562 | 6 | 0.754 | 0.429 | 4.940 |
| 9 | *Simplifying Life* | 403 | 1385 | 6.873 | 7 | 0.674 | 0.492 | 4.330 |

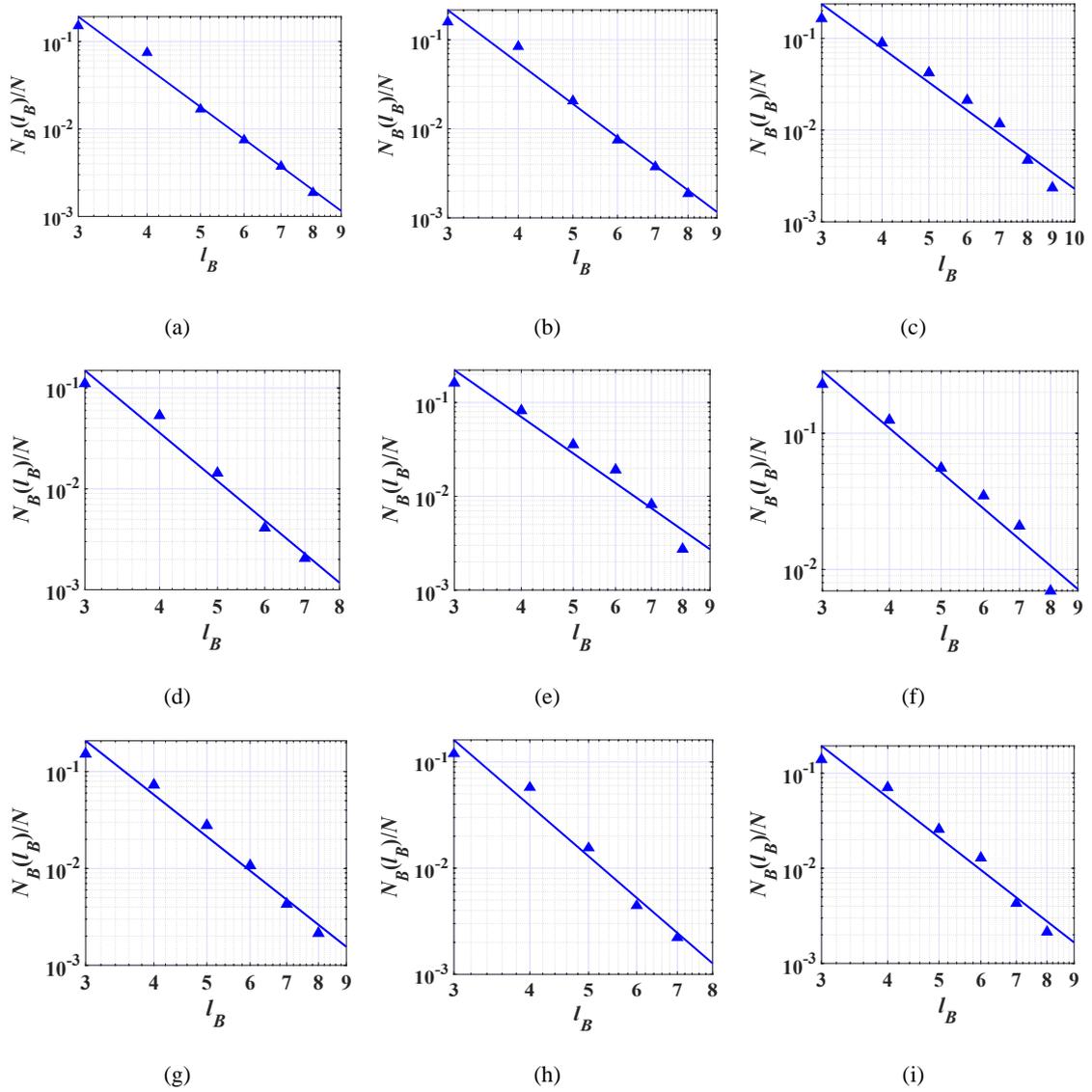

Fig. 3 The power-law relation between $N_B$ and $\ell_B$ of keyword co-occurrence networks, (a) *The Home Letter from Zeng Guofan*, (b) *No Longer Human*, (c) *Six Records of a Floating Life*, (d) *Decisive Moments in History*, (e) *A Brief History of Humankind*, (f) *The Kite Runner*, (g) *The Moon and Six Pence*, (h) *The Three-Body Problem*, (i) *Simplifying Life*.

4.2. Keyword Co-occurrence Network Fractal Characteristics Identification

Since the computation results for all 18 books are too verbose, the review text data from nine books were randomly selected to identify the fractal characteristics. The data were preprocessed, and a custom stop vocabulary was built to remove high-frequency meaningless words such as

'book', 'reading', and 'description'. Then, a keyword co-occurrence network was constructed for each book review using the methods described in Secs. 2.1 and 2.2. Taking the edge weight threshold $\alpha = 2$, the network was reduced, and the weight of the remaining edges was set to 1. An undirected and unweighted keyword co-occurrence network with an appropriate scale for each book was gained. Table 2 shows the typical topological features of those networks. Where N is the number of nodes, E is the number of edges, <k> is the average degree, D is the network diameter, C is the clustering coefficient, M is the modularity of the network and $d_B$ is the fractal dimension of the network. The sampling-based box-covering algorithm described in Sec. 2.3 was used on each network and the results are shown in Fig. 3. It can be seen that $N_B$ had a power-law relationship with $\ell_B$. The obtained fractal dimensions are shown in the last column of Table 2. Therefore, the keyword co-occurrence networks show fractal characteristics.

4.3. Topic Hierarchy Identification

Five keyword co-occurrence networks were selected from the nine networks in Section 4.2 for follow-up experiments to ensure that each of the five categories has a book's review text data used for all experiments. Applying the MCR algorithm to these five keyword co-occurrence networks, the fine topic hierarchies of review texts appeared, as shown in Figs. 4 and 5. Figure. 4(a) shows the topology of the network modules corresponding to the topics at each level in renormalizing the keyword co-occurrence network of *The Home Letter from Zeng Guofan*. It can be seen that the topological structures of the network modules at the bottom were all cliques, which is the same as the box representation in the box-sampling algorithm. Figure. 4(b) shows the fine hierarchical structure of the review topics in this book, which is divided into four layers. The first layer was the root of the hierarchical structure, which is generally summarised by the book name 'The Home Letter from Zeng Guofan'. The second layer expanded the root topic into two topics of 'Book Quality' and 'Evaluation Experience'. The third layer was divided into eight topics, which are extensions of the second layer topic. The coarse-grained topics on the third layer, such as 'Form', 'Content' and 'Insight', further subdivided the topic of 'Evaluation experience' on the second layer, reflecting the overall characteristics of products and services. The fourth layer was the most refined topic, such as 'Framework', 'Version', 'Values' and other fine-grained topics on the fourth layer subdivided the coarse-grained topics on the third layer, reflecting the evaluation of specific attributes. The topics were refined and supplemented layer by layer to meet various users' comprehensive and fine-grained needs for review understanding.

Figure 5 shows the fine-grained hierarchies of the book review topics of *No Longer Human*, *Six Records of a Floating Life*, *Decisive Moments in History* and *A Brief History of Humankind*. Those topic hierarchies reflect the focus of consumers on products and services, which is convenient for users to find the important information they need accurately and quickly, alleviating information overload. For example, *Six Records of a Floating Life* takes the life of the author and his wife as the mainline, narrating the ordinary but interesting home life and what they saw and heard when they travelled around. These narratives were fully displayed under the 'Description content' topic in Fig. 5(b). In *A Brief History of Humankind*, in addition to history, the author is also keen to make analytical descriptions from the perspectives of physics, chemistry, ecology and other disciplines. Therefore, under the topic of 'Theoretical knowledge' in Fig. 5(d), it included evolutionary history and knowledge elaboration of scientific disciplines such as physics and biology.

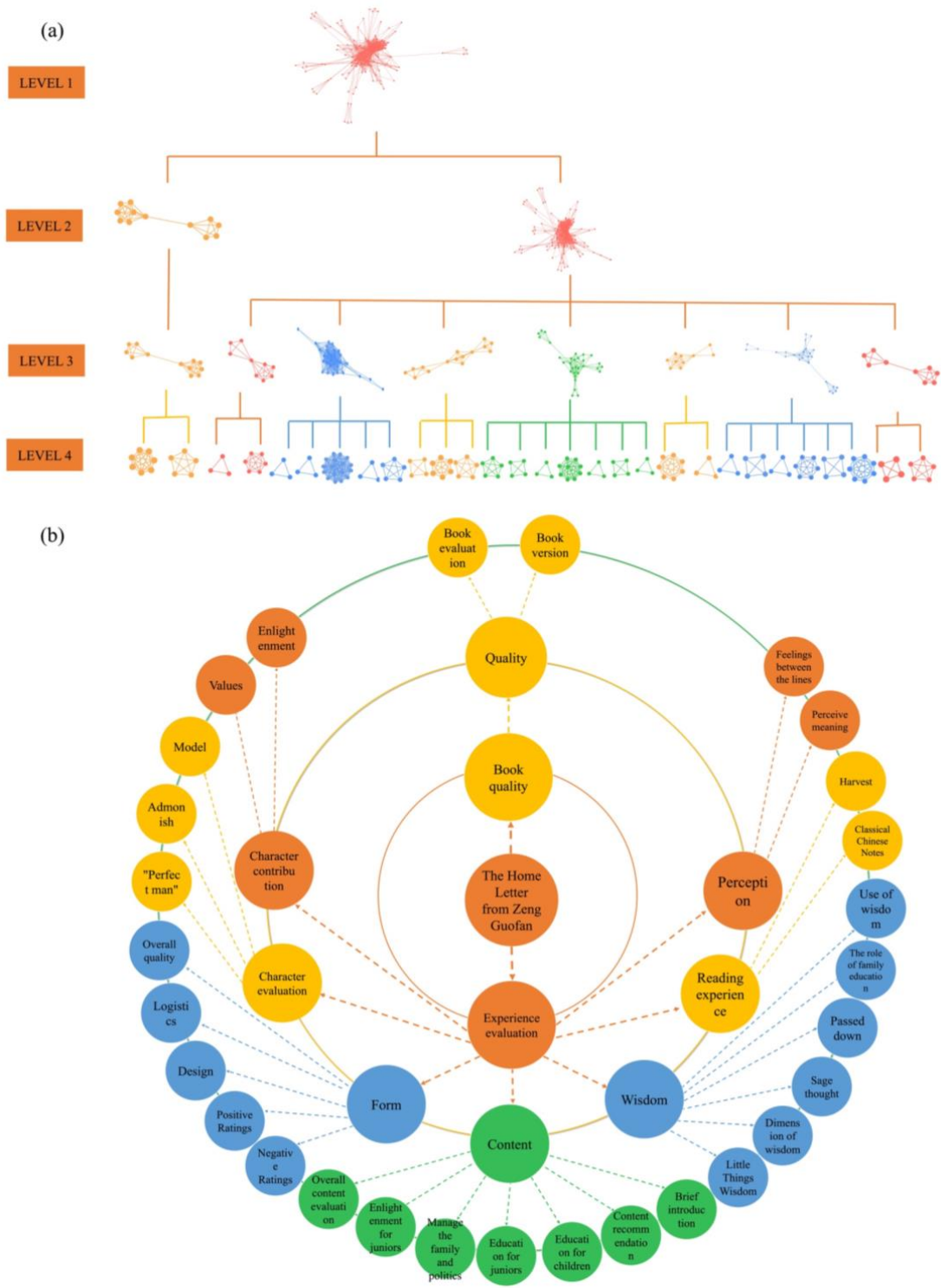

Fig. 4 Topic network module and hierarchy of *The Home Letter from Zeng Guofan*, (a) Network modules corresponding to different levels of topics, (b) The topic hierarchy of reviews of *The Home Letter from Zeng Guofan*.

(a)

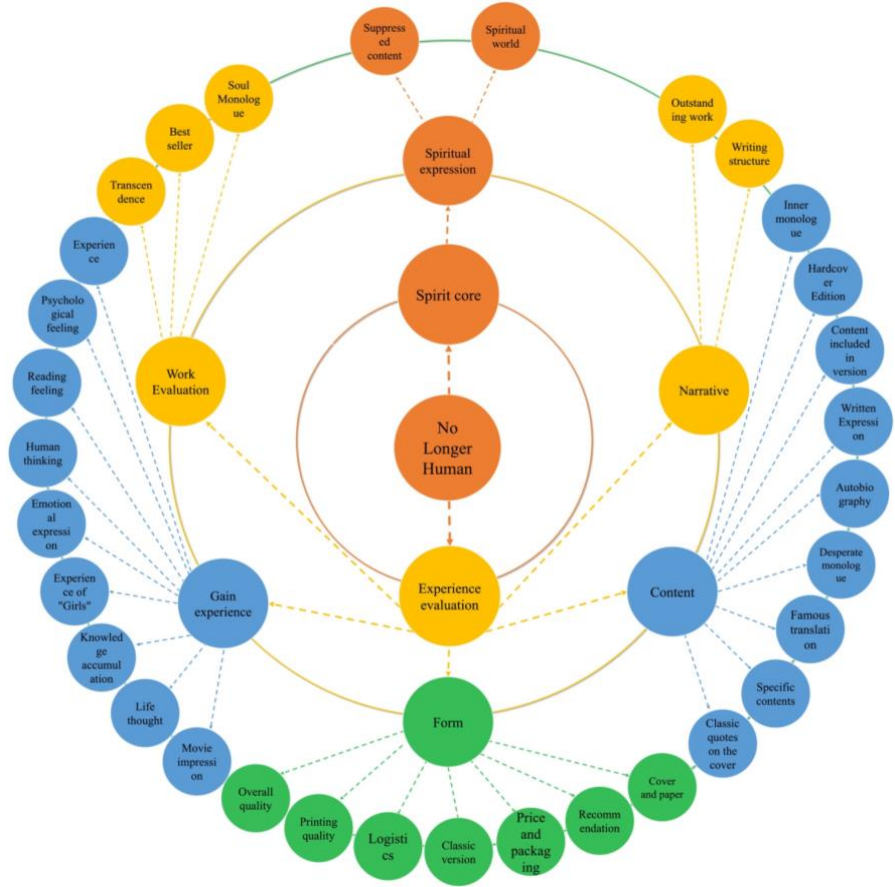

(b)

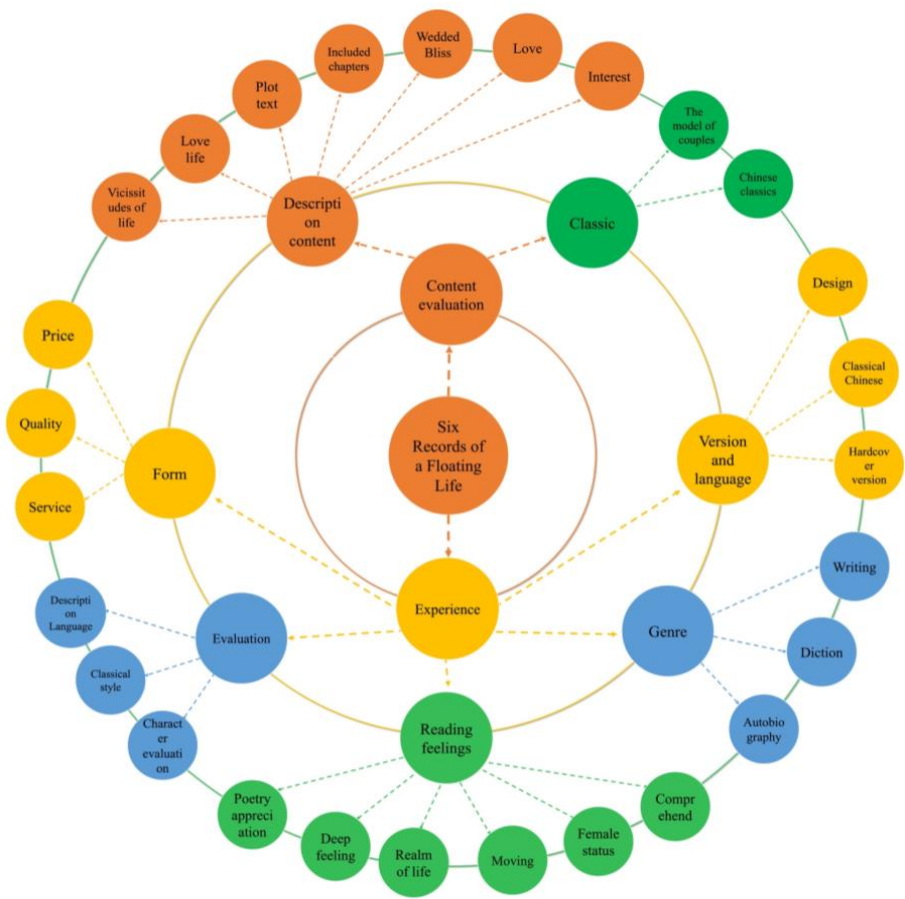

(c)

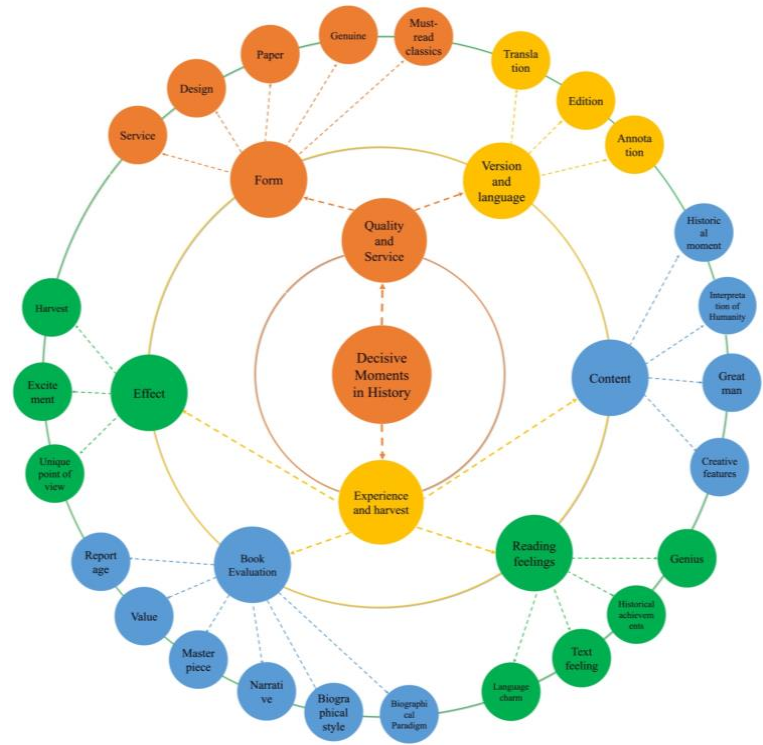

(d)

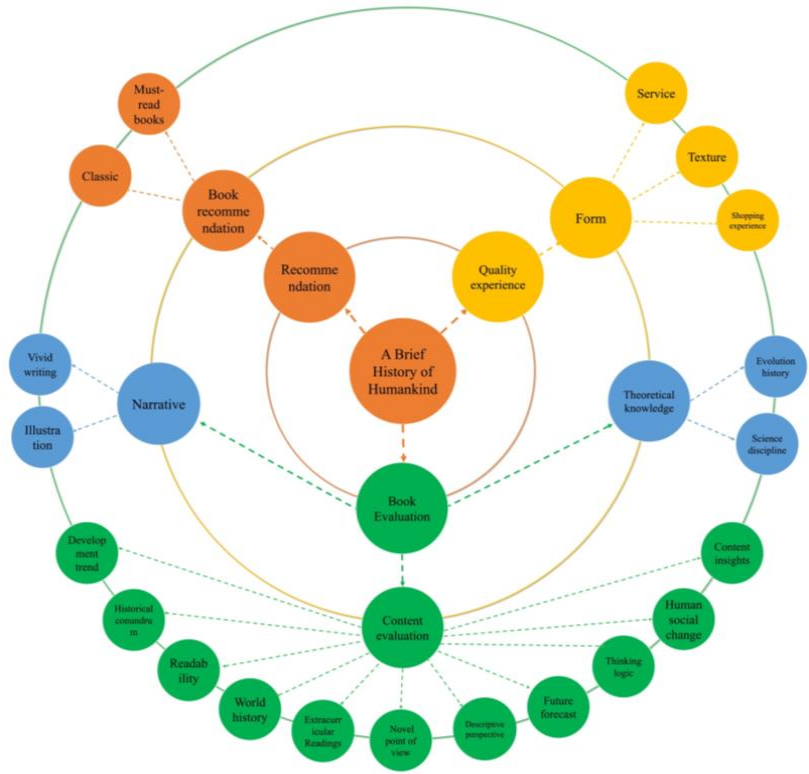

Fig. 5 Topic hierarchies of typical book reviews, (a) *No Longer Human*, (b) *Six Records of a Floating Life*, (c) *Decisive Moments in History*, (d) *A Brief History of Humankind*.

4.4. Review Text Classification

The method described in Sec. 3.3 was used for five selected books to achieve the classification of review texts. Some classification results are shown in Tables 3–7. The 'topic' column is part of the fine underlying topic corresponding to the book review text. The 'comment' column is a part of the review texts belonging to the topic. The 'probability' column is the possibility that the review text belongs to the corresponding topic. The case of probability 1 means that the review text belongs entirely to the corresponding topic. This is because the review text is short and the topic content it reflects is concentrated, and the keywords extracted from it belong to only one topic. Therefore, the probability that the review text belongs to the topic is calculated according to Eqs. (3)–(4) is 1. As shown in Tables 3–7, reviews obtained under each topic were highly related to the topic, and the classification effect was good. For example, topic 1 in Table 3 was 'Shopping experience', and the reviews classified under this topic were related to the buyer's feelings after purchasing the book. Topic 1 in Table 4 was 'Quality', and the reviews under this topic were all subjective or objective descriptions of book quality. Topic 2 in Table 5 was 'Writing', and the reviews under this topic were all descriptions and evaluations of the author's writing. Topic 1 in Table 6 was 'Biographical Paradigm', and the comments under this topic were all buyers' recognition of *Decisive Moments in History*, which is considered a biographical writing paradigm. Topic 10 in Table 7 was 'Design', and the reviews under this topic were all descriptions and evaluations of book bindings and covers.

Table 3 Partial Classification Results of *The Home Letter from Zeng Guofan*.

| Topics | Probabilities | Reviews |
|---|---|---|
| Topic 1 Shopping Experience | 0.802 | The paper of the book is good. We bought a lot together. It's cost-effective to catch up with the activity. |
|  | 0.814 | Thank Dangdang.com for providing us with such a good reading platform, and |

| | 0.920 | Buy books on Dangdang.com, and always trust Dangdang.com! The quality of the book is very good and it is worth reading! |
| | 0.823 | Genuine good books and considerate service. Since then, buy books and come to Dangdang.com! |
| | 0.873 | Good stuff, great value for money, reasonable price, high-cost performance! |
| | 0.924 | This book is cost-effective and detailed. I like it! very nice! Worth recommending! |
| Topic 2 Dimension of Wisdom | 0.693 | Zeng Guofan's family letter is a broad and profound University question for self-cultivation, scholarship, and making friends. |
| | 0.485 | Zeng Guofan's family letters are very good, with good expositions on scholarship, family harmony, life, making friends, and so on. |
| | 1.000 | Learn Zeng Guofan for politicians and Rockefeller for business. |
| | 0.493 | 1300 letters repeatedly told brothers and children the 'basic methods' of 'reading, making friends, self-cultivation, and maintaining a family'! |
| | 0.625 | Zeng Guofan's family letter is classic, education, life, financial management, well, I haven't finished reading it yet. |
| | 0.741 | His life journey includes self-cultivation, family harmony, politics, scholarship, making friends, and employing people. |
| Topic 5 Education for Juniors | 0.732 | Personal experience is an article, and the letter contains Zeng Guofan's teaching his children the way to live in the world, and the Tao is in it! |
| | 0.803 | Look at how the predecessor educates their children. It's okay. |
| | 0.547 | Zeng Guofan has always been my idol. No matter educating children or doing things personally, I have benefited a lot from reading it! |
| | 1.000 | It is thought-provoking to teach your brother from the perspective of a brother! |
| | 1.000 | In his family instruction, Zeng Guofan taught his children according to their aptitude. |
| | 1.000 | He taught his younger brother and children according to their aptitude. |

Table 4 Partial Classification Results of *No Longer Human*.

| Topics | Probabilities | Reviews |
| --- | --- | --- |
| Topic 1 Quality | 1.000 | Haven't started to look at it yet, but there are no quality problems caused by shipping packaging problems. |
| | 0.874 | This book is very good, comprehensive in content, beautiful pictures, worth buying. |
| | 1.000 | The cover of the book feels great. The texture of the suede and the design of the inner tattoo is also very good-looking. |
| | 1.000 | The quality of the printed text is very good, and it can be spread out evenly. |
| | 0.905 | The book is very good, the packaging, printing, paper quality, content, etc. are all very good. |
| | 1.000 | The printing is clear and the binding is beautiful. It is worth reading. |
| Topic 2 Human | 0.445 | The ugliness and hypocrisy of human nature are invisible! The delicate and sentimental text starts here. |

| Topics | Probabilities | Reviews |
| --- | --- | --- |
| Thinking | 1.000 | This is human disqualification, where the ugliness and hypocrisy of human nature are invisible. |
| | 0.803 | The protagonist in the book seems to be the epitome of the dark side of everyone's heart, and his thoughts on human nature are flogged. |
| | 0.633 | I think the *No longer human* is written in the depths of human nature, which is worthy of our exploration. |
| | 0.873 | The profound understanding of human nature is more analysis. |
| | 1.000 | Human nature is the most complicated thing in the world. |
| Topic 21 Must-read Classics | 0.942 | Must-read classics, warm people. |
| | 0.837 | Recommended by the school's required reading list, but not as a task. Rejoicing and harvesting. |
| | 0.815 | This book is a must-read classic and fascinating. |
| | 0.858 | Must-read classics, fascinating, clear printing, reasonable layout, reading, and eye protection, very good! |
| | 0.820 | Classic, good plot. |
| | 0.593 | *No longer human*, beautifully decorated, must-read classics, highly recommended. |

Table 5 Partial Classification Results of *Six Records of a Floating Life*.

| Topics | Probabilities | Reviews |
| --- | --- | --- |
| Topic 1 Design | 0.603 | The packaging is beautiful, the content is substantial, and the real thing matches the picture. |
| | 1.000 | This is by far my favorite version I've found, the translation is great and detailed. |
| | 0.591 | I like packaging very much. The paper is also very textured and the pattern is exquisite. |
| | 1.000 | I really like the layout of the original text, annotations, and translations. |
| | 0.549 | The packaging is beautiful and the content layout is also very practical. |
| | 1.000 | The frame is well designed. |
| Topic 2 Writing | 0.556 | The ancient diary writing is comfortable. |
| | 0.759 | I still believe more in Mr. Shen Fu's writing. |
| | 0.519 | The writing is delicate and the emotion is sincere. |
| | 1.000 | The content is good and the writing is good. |
| | 0.833 | The writing is simple and plain, and the emotion is sincere. |
| | 0.779 | The writing is beautiful, and the author translated the charm. |
| Topic 9 Service | 0.578 | I'm very satisfied with the book. I like the pure Chinese style of ink and wash. The delivery speed is also great. |
| | 1.000 | Fast delivery and good service. |
| | 1.000 | Well packaged, full of dry goods. |
| | 0.844 | The express delivery has been received, and the outer package is very good without damage. |
| | 0.806 | The package is intact and the logistics is fast. We look forward to the content. |
| | 1.000 | Fast delivery, complete packaging, satisfied. |

Table 6 Partial Classification Results of *Decisive Moments in History*.

| Topics | Probabilities | Reviews |
|---|---|---|
| Topic 1 Biographical Paradigm | 1.000 | A model of biography, a good book is worth recommending. |
| | 0.474 | After reading it, I am not very interested in world history, but it is quite interesting as a biography of a character. |
| | 1.000 | It's a great example of biographical writing. |
| | 0.527 | This book tells about twelve people who have influenced the course of human history, unlike ordinary biographies. |
| | 1.000 | One of the representative works of the king of biography. |
| | 1.000 | A classic of biographical literature. |
| Topic 2 Service | 1.000 | The book is not a hardcover edition, but both paper and printing are good. Recently, I'm obsessed with European history. |
| | 0.644 | The book has been received, the logistics is full, and the book is not bumped. |
| | 0.651 | Beautifully printed, rich in content, highly readable, great price, great logistics. |
| | 1.000 | The packaging is good and the book print is clear. |
| | 0.734 | The paper is thin, the printing is clear, and the outer packaging is not damaged. |
| | 1.000 | The book should be genuine and well packaged. |
| Topic 23 Classic | 0.751 | Beautifully bound, classic must read, beautiful and meaningful. |
| | 0.899 | Recommended by the teacher, it is a classic worth reading. |
| | 0.473 | Classic must-read books, love the cover design. |
| | 0.741 | A must-read classic. |
| | 1.000 | A must-read classic, the condensed history of mankind. |
| | 1.000 | A must-read classic series. |

Table 7 Partial Classification Results of *A Brief History of Humankind*.

| Topics | Probabilities | Reviews |
|---|---|---|
| Topic 5 Service | 1.000 | Beautifully packaged and affordable. |
| | 0.947 | Logistics is fast, the book is well packaged and the content is beautiful. |
| | 0.946 | The packaging is very good and the logistics is fast. Let's see the content again. |
| | 0.889 | Well packaged, good content |
| | 0.791 | The packaging is ok. I haven't read it yet. I'm looking forward to it |
| | 1.000 | Genuine books, fast delivery, full marks! |
| Topic 7 Language and Writing | 0.867 | The writing is more vivid, and the angle of entry is different from that of general history books, which is very novel. |
| | 0.556 | A very wonderful book, the author writes the history of mankind in vivid words. |
| | 0.703 | The language is very vivid and interesting. Many views coincide with me and are worth seeing. |
| | 0.703 | The language is vivid, and a brief history is told in brisk language, which makes people fall in love with it. |
| | 0.916 | A very humorous history book, vivid, professional and imaginative. |
| | 0.703 | The book is great, the language is vivid and worth having! |
| Topic 10 | 0.875 | The paper and printing are first-class. The disadvantage is that I don't like this |

| Design | | kind of white pages. |
|---|---|---|
| | 0.908 | A book worth reading! From the paper to the content is super awesome! |
| | 0.506 | The printing quality is excellent, and the packaging is very novel. I like it very much. |
| | 0.821 | The design of the packaging and cover paper is so thoughtful! |
| | 1.000 | The quality is very good and the pictures are very rich! |
| | 1.000 | The printing quality is good. |

## 5. ALGORITHM PERFORMANCE EVALUATION

5.1. Evaluation Methods and Comparison Algorithms

The method proposed in this study was compared with commonly used methods from the obtained topic hierarchy, text classification effect and algorithm time performance to verify its effectiveness.

The topic hierarchy analysis mainly evaluates the topic hierarchy rationality, the topic coherence, and the topic independence obtained by the MCR algorithm. The time performance analysis mainly evaluates the time required for the MCR algorithm to obtain the topic hierarchy on the sample dataset. The HLDA algorithm and Louvain algorithm were used as comparison methods. The HLDA algorithm is a typical topic hierarchy mining method suitable for reviews. The parameter setting of the HLDA referred to the modelling strategy proposed in reference 43. The specific parameters were set as follows: $L = 4$, $\eta = 1$, $\gamma = 10$. Since the MCR algorithm identifies the topic hierarchy based on the keyword co-occurrence network, the hierarchical community mining algorithm in complex networks is also suitable for this scenario. As a typical hierarchical community mining algorithm, the Louvain algorithm has superior time performance and a wide range of applications. Therefore, this study chose the HLDA and Louvain algorithms for comparison.

In the evaluation of review text classification effects, in addition to comparing the text classification results obtained in Sec. 4.4 with the HLDA and Louvain algorithms, this study also selected the LDA topic model and the TextCNN algorithm based on the word vector[44]. Although the LDA and TextCNN algorithms are not hierarchical topic classification methods, they are the most commonly used text classification methods. The number of LDA topics was determined according to the smallest perplexity. The word vector was obtained from training on experimental data using CBOW, and then a TextCNN feature fusion model was built based on the TensorFlow deep learning framework.

5.2. Topic Hierarchy Rationality

A reasonable topic hierarchy means that topics closed to the root node have more general semantics, while those close to leaf nodes have more specific semantics.[45] Based on this general principle, Kim et al. proposed the Topic Specialisation $TS(z_i)$, which measures whether the topic hierarchy is reasonable by calculating the semantic distance between each layer of topics and the general topic $z_{norm}$.[46]

$z_{norm}$ is defined as the topic with the most general semantics, representing the word distribution in the entire keyword set $V$. The probability $p(v_i|z_{norm})$ that the keyword $v_i$ appears in $z_i$ is proportional to the frequency $D(v_i)$ of the document containing the keyword $v_i$, and the calculation equation is shown in Eq. (8). Where $\beta$ refers to the smoothing factor. The Topic Specialisation value $TS(z_i)$ is obtained by calculating the deviation between $z_i$ and $z_{norm}$, as shown in Eq. (9). It can be seen that the deviation between $z_i$ and $z_{norm}$ has a negative correlation with the cosine similarity between the topics. The smaller the Topic Specialisation value of $z_i$, the more common semantics of $z_i$, and the position of $z_i$ closer to the root node.

$$p(v_i|z_{norm}) = \frac{D(v_i) + \beta}{\sum_{j=1}^{n} D(v_j) + \beta |V|} \tag{8}$$

$$TS(z_i) = 1 - \frac{z_i \cdot z_{norm}}{|z_i| \cdot |z_{norm}|} \tag{9}$$

$$TS_{avg} = \frac{sum(TS(z_i))}{I}(i = 1, 2, \cdots, I) \tag{10}$$

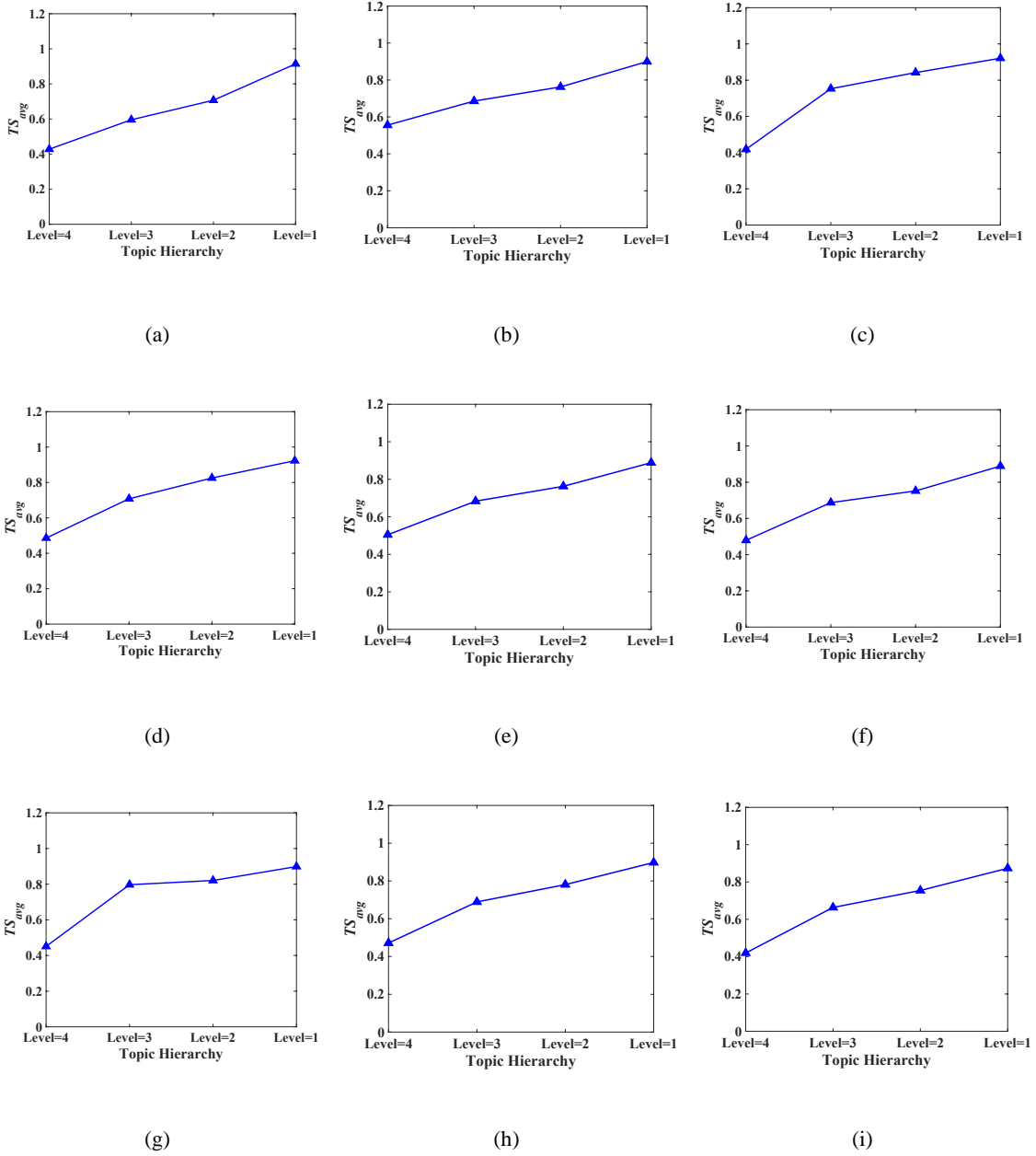

Fig. 6 Hierarchy analysis results, (a) *The Home Letter from Zeng Guofan*, (b) *No Longer Human*, (c) *Six Records of a Floating Life*, (d) *Decisive Moments in History*, (e) *A Brief History of Humankind*, (f) *The Kite Runner*, (g) *The Moon and Six Pence*, (h) *The Three-Body Problem*, (i) *Simplifying Life*.

The Average Topic Specialisation value $TS_{avg}$ for topics at each layer is calculated in Eq. (10). As shown in Fig. 6, the $TS_{avg}$ of each topic hierarchy of all nine book datasets increased from the root node to the leaf node, that is, the topic near the root node had a small deviation from $z_{norm}$, and more general expression semantics, while the topic close to the leaf node had a large deviation from $z_{norm}$ and more specific expression semantics. Therefore, the topic hierarchy identified by MCR is reasonable and effective.

5.3. Topic Coherence

Topic Coherence $C(z_i)$ is used to evaluate the quality of topics in a given topic hierarchy.[47] Topic Coherence is measured according to the co-occurrence relationship of keyword pairs under the topic. The more keyword pairs co-exist under the topic, the better the cohesion within the topic and the more coherent the semantics expressed. Compared with perplexity, Topic Coherence is consistent with human interpretation and can correctly judge whether a topic has coherent semantics.[48] $C(z_i)$ and the Average Topic Coherence values $C_{avg}$ are calculated as follows:

$$C(z_i) = \sum_{m=2}^{M} \sum_{l=1}^{m-1} \log \frac{D(v_m^{z_i}, v_l^{z_i}) + 1}{D(v_l^{z_i})} \tag{11}$$

$$C_{avg} = \frac{sum(C(z_i))}{I} (i = 1, 2, \cdots, I) \tag{12}$$

Where $z_i = (v_1^{z_i}, v_2^{z_i}, \cdots, v_M^{z_i})$ represents the topic $z_i$, consisting of the top M words with the largest probabilities; $D(v_m^{z_i}, v_l^{z_i})$ represents the document frequency of co-occurrence of word $v_m^{z_i}$ and word $v_l^{z_i}$; $D(v_l^{z_i})$ represents the document frequency containing the word $v_l^{z_i}$.

The Average Topic Coherence values $C_{avg}$ of the bottom-level topics in the topic hierarchies of nine book reviews were obtained by the MCR, HLDA, and Louvain algorithms. As shown in Fig. 7, MCR achieved the highest $C_{avg}$ in all nine datasets, which indicates that the topics obtained by MCR have more coherent semantics and are more in line with the human interpretation.

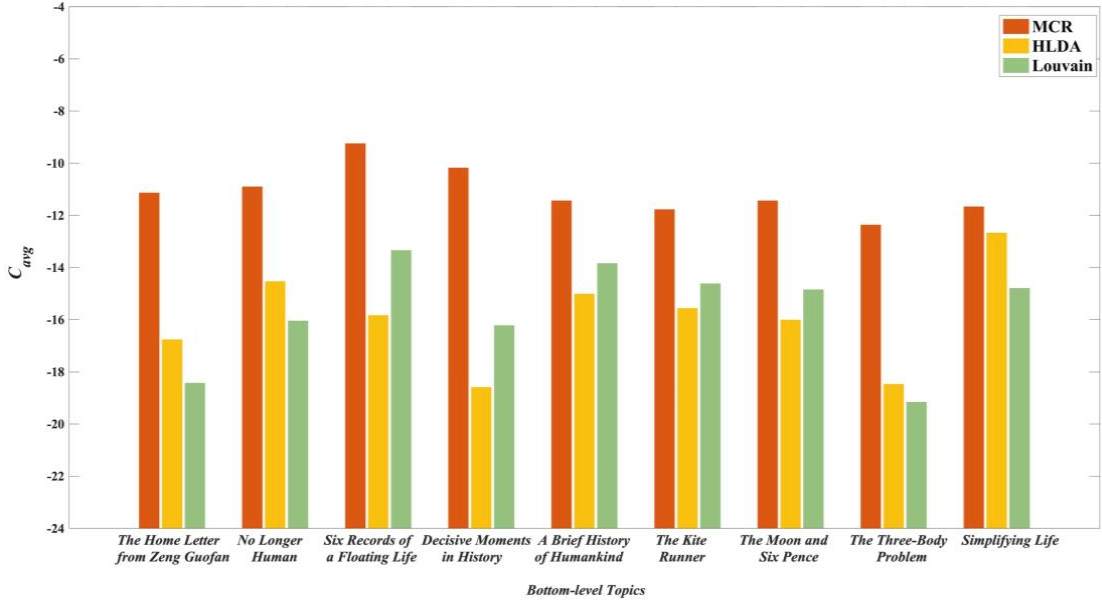

Fig. 7 The result of Topic Coherence comparison.

5.4. Topic Independence

Topic Independence $I_{avg}$ is used to evaluate the discrimination between topics.[49] Comparing the independence between topics is essentially comparing whether there are differences in keywords under the topics. High independence between the topics means that the differences between the topics are large, the coupling degree is low and the clustering effect is better. $I_{avg}$ is calculated as follows：

$$S_{ij} = \frac{V_i \cdot V_j}{|V_i||V_j|} \tag{13}$$

$$I_{avg} = 1 - \frac{2\sum_{i=1}^{K}\sum_{j=1,j\neq i}^{K} S_{ij}}{K(K-1)} \tag{14}$$

Where $K$ is the number of topics in a layer; topic $z_i$ is represented by a feature vector $V_i = (v_1^i, v_2^i, \cdots, v_n^i)$; $n$ is the number of unique keywords contained in all topics; $v_a^i$ is the distribution frequency of the $a$-th keyword in $z_i$, if $z_i$ does not contain the $a$-th keyword, then $v_a^i = 0$; $S_{ij}$ represents the similarity between the $z_i$ and the $z_j$.

$I_{avg}$ of the bottom-level topics in the topic hierarchies of nine book reviews were obtained by the MCR, HLDA and Louvain algorithms. As shown in Fig. 8, topic hierarchies identified by MCR and Louvain algorithms were 1.00 since there was no repetition of keywords between topics in the same layer, which was stronger than HLDA. Therefore, the topic hierarchy identified by MCR has high independence and good discrimination.

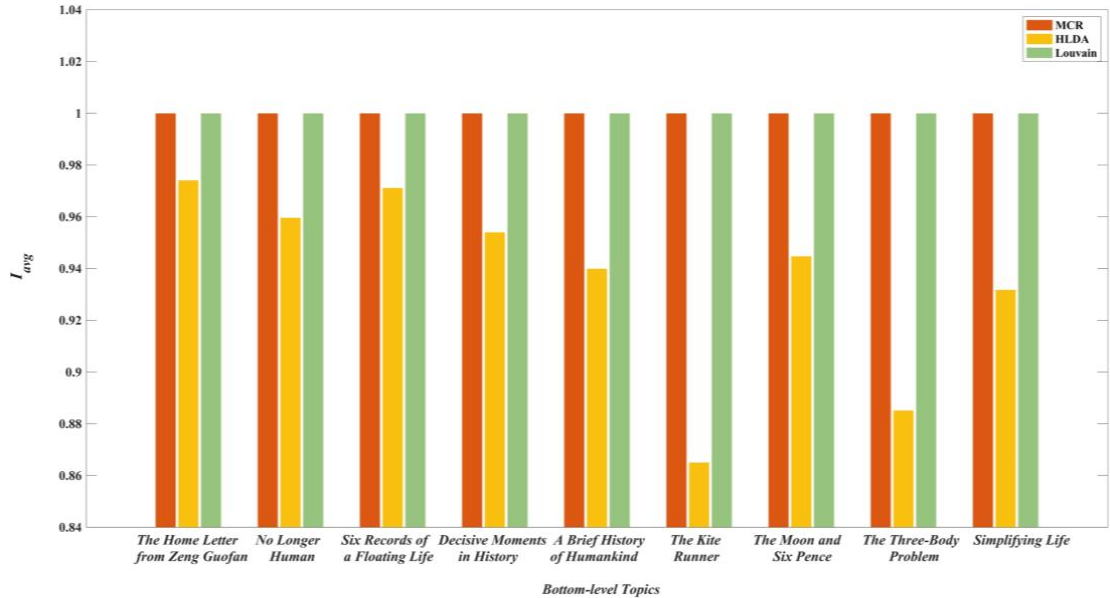

Fig. 8 The result of Topic Independence comparison.

5.5. Time Performance

To further evaluate the performance of the MCR algorithm, we compared the average running time of the MCR, HLDA and Louvain algorithms. The computer used was with the following configuration: (i) Hardware: Intel(R) Core(TM) i7-7700 HQ processor, 8.0 GB memory; (ii) Operating system: Windows10; (iii) Computing environment: Python 3.8.5.

Table 8 shows the average running time (seconds) to get topic hierarchies of nine book review texts. The time performance of the Louvain algorithm is the best since its complexity is linear on typical and sparse data sets. The HLDA is the slowest, which costs most of the time on Gibbs sampling. The average running time of the MCR and Louvain algorithm can both get the topic hierarchy within an acceptable time elapse.

Table 8 The Average Running Time.

| Datasets | MCR/sec | HLDA/sec | Louvain/sec |
|---|---|---|---|
| *The Home Letter from Zeng Guofan* | 2.870 | 192817.436 | 2.277 |
| *No Longer Human* | 2.351 | 190322.239 | 1.608 |
| *Six Records of a Floating Life* | 1.514 | 188629.060 | 1.505 |
| *Decisive Moments in History* | 1.581 | 208749.720 | 1.504 |
| *A Brief History of Humankind* | 0.691 | 186111.306 | 0.516 |
| *The Kite Runner* | 0.109 | 42620.171 | 0.521 |
| *The Moon and Six Pence* | 1.239 | 171134.724 | 1.050 |
| *The Three-Body Problem* | 0.910 | 198722.192 | 0.802 |
| *Simplifying Life* | 0.868 | 92755.821 | 1.117 |

5.6. Text Classification Result

5.6.1. Evaluation index

The topic hierarchy ultimately serves the classification of Internet review texts. Evaluation indexes for text classification are accuracy rate (P), recall rate (R) and F1 value. According to the classification results, the confusion matrix[50] is established as shown in Table 9:

(1) The accuracy rate P refers to the ratio of the number of samples correctly identified as the class to the total number of samples identified as the class. This reflects the accuracy of the classification results. The accuracy rate is defined as follows：

$$P = \frac{TP}{TP + FP} \tag{15}$$

(2) The recall rate R refers to the ratio of the number of samples correctly identified as the class to the total number of samples that belong to the class. This reflects the completeness of the classification results. The recall rate is defined as follows:

$$R = \frac{TP}{TP + FN} \quad (16)$$

(3) To fully reflect the classification performance, the comprehensive index F-measure value was also defined as follows:

$$F_\beta = \frac{(\beta^2 + 1) \times P \times R}{\beta^2 \times P \times R} \quad (17)$$

where $\beta$ is the adjustment parameter used to adjust the ratio of the accuracy rate P and the recall rate R in the calculation formula. In this study, $\beta = 1$ was taken to obtain the F1 value.

Table 9 Confusion Matrix.

| Actual situation | Discriminant situation | |
|---|---|---|
| | Discriminant positive case | Discriminant negative case |
| Actual positive case | True-positive (TP) | False-negative (FN) |
| Actual negative case | False-positive (FP) | True-negative (TN) |

5.6.2. Evaluation results

The performance of the text classification method in this study and four comparison methods for review text classification were compared. The accuracy rate, recall rate and F1 value of the classification results in Sec.5.6.1 were calculated. As shown in Table 10, the classification method in this study performed the best in terms of classification accuracy rate, recall rate and F1 value, followed by the Louvain-based classification method and TextCNN methods, and the HLDA-based classification method and LDA performed the worst. Compared with the four comparison methods, in terms of accuracy rate, the method in this study improved the highest by 18.481% and the lowest by 14.649%. In terms of recall rate, the method in this study improved the highest by 19.747% and the lowest by 12.720%. In terms of the F1 value, the method in this study improved the highest by 17.796% and the lowest by 14.205%. This indicates that the classification method in this study is effective and feasible for a short review text classification.

Table 10 The Result of Comparative Experiment.

| Datasets | Methods | Accuracy rate (P)/% | Recall rate (R)/% | F1-value (F1)/% |
|---|---|---|---|---|
| Reviews of *The Home Letter from Zeng Guofan* | MCR-based classification method | 91.830 | 93.487 | 92.651 |
| | HLDA-based classification method | 77.935 | 80.972 | 79.424 |
| | Louvain-based classification method | 84.315 | 83.507 | 83.909 |
| | LDA | 77.181 | 79.753 | 78.446 |
| | TextCNN | 86.164 | 87.292 | 86.724 |
| Reviews of *No Longer Human* | MCR-based classification method | 87.274 | 91.935 | 89.544 |
| | HLDA-based classification method | 71.481 | 75.285 | 73.333 |
| | Louvain-based classification method | 85.594 | 86.081 | 85.836 |
| | LDA | 76.503 | 80.452 | 78.428 |
| | TextCNN | 78.152 | 81.574 | 79.826 |
| Reviews of *Six Records of a Floating Life* | MCR-based classification method | 88.167 | 89.108 | 88.635 |
| | HLDA-based classification method | 76.772 | 84.771 | 80.573 |
| | Louvain-based classification method | 85.661 | 86.587 | 86.122 |
| | LDA | 69.686 | 73.623 | 71.600 |
| | TextCNN | 83.016 | 84.049 | 83.529 |
| Reviews of *Decisive Moments in History* | MCR-based classification method | 89.300 | 90.397 | 89.846 |
| | HLDA-based classification method | 73.506 | 70.650 | 72.050 |
| | Louvain-based classification method | 78.436 | 80.091 | 79.254 |
| | LDA | 73.836 | 79.817 | 76.711 |
| | TextCNN | 80.214 | 82.913 | 81.541 |
| Reviews of *A Brief History of Humankind* | MCR-based classification method | 88.543 | 91.185 | 89.845 |
| | HLDA-based classification method | 77.114 | 81.629 | 79.307 |
| | Louvain-based classification method | 82.524 | 84.165 | 83.337 |

| | | | |
|---|---|---|---|
| LDA | 72.498 | 78.465 | 75.364 |
| TextCNN | 82.193 | 83.424 | 82.804 |

## 6. CONCLUSIONS

Mining the hierarchical structure of review topics, realising the fine classification of review texts, and meeting users' requirements for a comprehensive and accurate grasp of review texts can help alleviate information overload. However, existing hierarchical topic classification methods rely on external corpora and human intervention. Therefore, in Internet review texts ,this study proposes a topic hierarchy mining method using MCR. The following are the most innovative contributions:

(1) The self-similarity structure in the keyword co-occurrence network was revealed for the first time, showing fractal characteristics in the review text topics on the Internet. Based on word segmentation and stop word removal, the TextRank algorithm was used to extract text keywords and constructed the keyword co-occurrence network. After sampling-based box-covering, the self-similarity structure in the keyword co-occurrence network was discovered, and the fractal characteristics existing in the review text were identified. It lays a foundation for the subsequent use of fractal network theory to identify the topic hierarchy of review texts.

(2) The MCR algorithm was constructed and proved its latent capacity to identify the topic hierarchy in review texts. Because intermediate frequency words reflect text features better in text mining, the edge adjacency entropy and the box distance were defined to modify the renormalisation procedure, and the MCR algorithm to identify the fine topic hierarchy was proposed. Analysing and comparing quantitative indicators, the topic hierarchy identified by MCR is reasonable and performs well in topic coherence and topic independence. MCR has obvious advantages in terms of time consumption over the HLDA algorithm.

(3) The bottom-level topic categories were extended based on MCR to achieve reliable text classification of short Internet reviews. Using the extended bottom-level topic categories, the classification of the review texts was completed according to the probabilistic topic model. Compared with the HLDA, Louvain, LDA, and TextCNN methods, the MCR-based method significantly improved the classification accuracy rate, recall rate, and F1 value.

The following aspects need to be further explored in this study: (i) Although using random sampling and greedy strategies, box-covering is still time-consuming and needs to be optimised for large-scale review text topic hierarchy identification. (ii) The potential effectiveness of the method proposed in this study for topic mining of more types of texts remains to be further verified.

## ACKNOWLEDGMENTS

This study is supported by the National Key Research & Development Plan Project (2021YFF0900200), National Natural Science Foundation of China (72002017), Youth Talent Promotion Program of Beijing Association for Science and Technology (20200016), Program for Promoting the Connotative Development of BISTU (5212110901, 5026010961) and Research Foundation Project of BISTU (2021XJJ42).

The authors are grateful to the anonymous reviewers and the editor for their valuable comments and suggestions that have greatly improved the quality of this study.